\documentclass[preprint,12pt]{elsarticle}

\usepackage{amssymb}
\usepackage{amsmath}

\journal{Chaos, Solitons and Fractals}

\begin{document}

\begin{frontmatter}



\title{Self-oscillation excitation under condition of positive dissipation in a state-dependent potential well}


\author{Vladimir V. Semenov}

\address{Department of Physics, Saratov State University, Astrakhanskaya str. 83, 410012 Saratov, Russia}

\begin{abstract}
The self-oscillatory dynamics is considered as motion of a particle in a potential field in the presence of dissipation. Described mechanism of self-oscillation excitation is not associated with peculiarities of a dissipation function, but results from properties of a potential, whose shape depends on a system state. Moreover, features of a potential function allow to realize the self-oscillation excitation in a case of the dissipation function being positive at each point of the phase space. The phenomenon is explored both numerically and experimentally on the example of a double-well oscillator with a state-dependent potential and dissipation. After that a simplified single-well model is studied.
\end{abstract}


\end{frontmatter}


\section{Introduction}
\label{intro}
The problem of the self-sustained oscillation existence in autonomous dynamical systems was originally introduced by H.Poincare \cite{poincare1881}. Thereafter bifurcation mechanisms of self-sustained oscillation excitation were described by A.Andronov and his associates \cite{andronov1966} and E.Hopf \cite{hopf1942,morawetz2002}. Despite the issue of the self-sustained oscillations is known for over a century, this topic is still attractive and interesting. This is due to the fact that self-sustained oscillators are of a frequent occurrence in physics \cite{groszkowski1964,hassard1981,jenkins2013,aguliar2015}, chemistry \cite{field1985,epstein1998,mikhailov2013}, geology \cite{costa2012,lauro2011}, climatology \cite{berger1984,cane1990,munnich1991}, biology \cite{selkov1968,adey1984,walleczek2000,izhikevich2007}, economics \cite{goodwin1949,mishchenko2014} and other fields. Problematics of self-oscillations in deterministic systems includes regular periodical motions as well as the chaotic dynamics \cite{strogatz1994,wiggins2003,anishchenko2007}. There are well-known stochastic effects, which are related to the self-oscillation topic: stochastic bifurcations (in the context of the stochastic Andronov-Hopf bifurcation \cite{ebeling1986,lefever1986,fronzoni1987,arnold2003,zakharova2011}), coherence resonance \cite{pikovsky1997,lindner2004,zakharova2010,geffert2014,semenov2015,masoliver2017}, stochastic synchronization \cite{pikovsky2001,goldobin2004,zakharova2013}, stochastic resonance \cite{yamapi2017,semenov2016}. Since a class of the self-oscillatory systems includes a broad variety of dynamical systems with different nature and features, the problem of general description and interpretation is significantly important. This issue will be discussed in the present paper.

The dynamics of an oscillator with one degree of freedom can be interpreted as motion of a particle in a potential field in the presence of dissipation. In that case a mathematical model of the oscillator is presented in the following form:

\begin{equation}
\label{general-eq}
\dfrac{d^{2}y}{dt^{2}}+\gamma\dfrac{dy}{dt}+\dfrac{dU}{dy}=0,
\end{equation}
where the factor $\gamma$ characterizes dissipation, $U$ is a function denoting the potential field. Typically, the potential function is assumed to be univariate: $U=U(y)$. Depending upon the specific of the dynamical system (\ref{general-eq}), the dissipation factor $\gamma$ can be either a fixed parameter, $\gamma=const$, or a state-dependent function $\gamma=\gamma(y,\frac{dy}{dt})$.  In the presence of constant dissipation ($\gamma=const$) the deterministic model (\ref{general-eq}) exhibits two  kinds of the behaviour in the potential well: either damped oscillation in case $\gamma>0$ or oscillations with unlimited amplitude growth in case $\gamma<0$. If the dissipation depends on a system state and the function $\gamma(y,\frac{dy}{dt})$ possesses negative values in some area of the phase space,  self-oscillation excitation can be achieved. In such a case the self-oscillation excitation is a result of the negative dissipation action, which is associated with pumping of energy.

\begin{figure}
\centering
\includegraphics[width=0.7\textwidth]{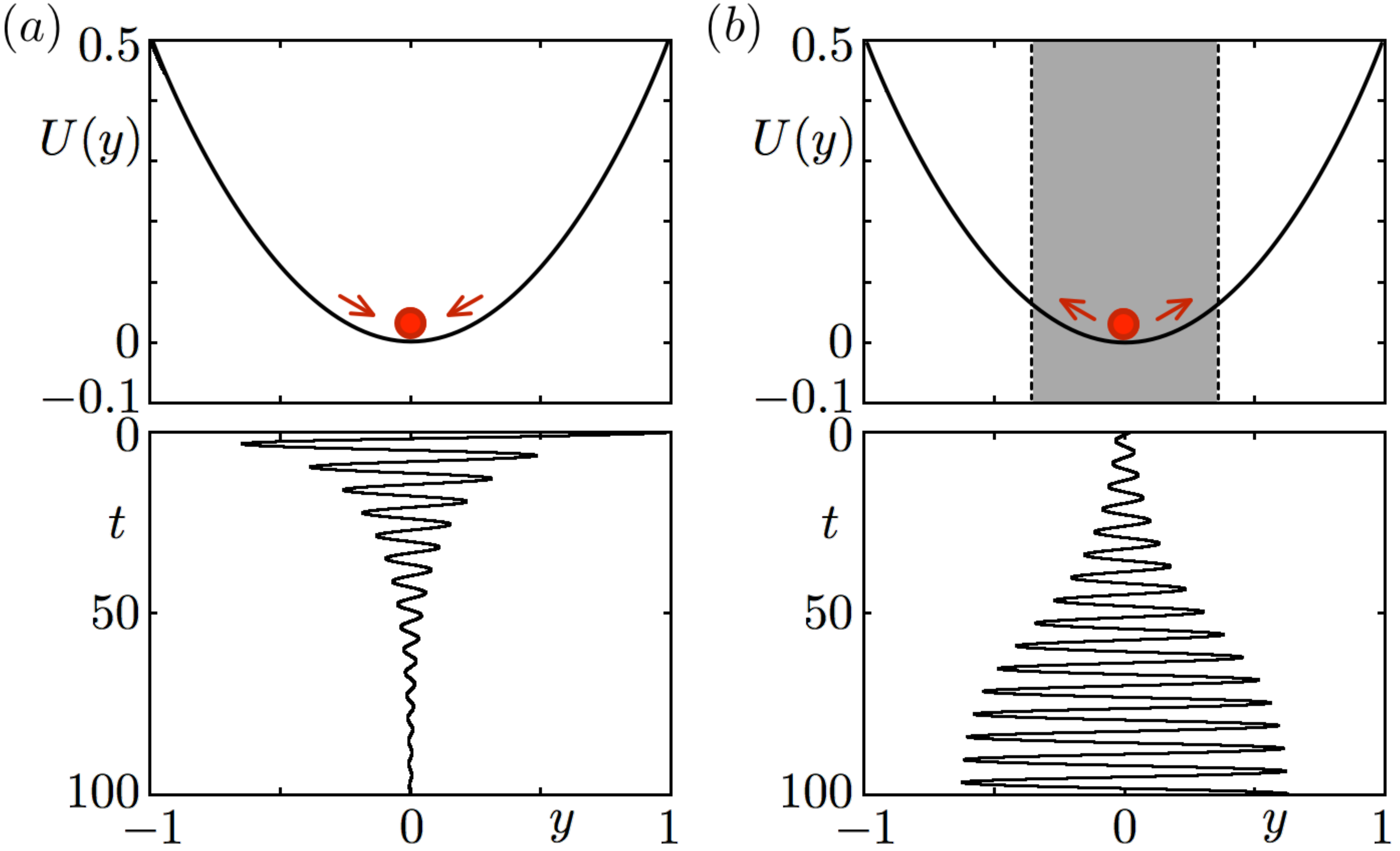} 
\caption{System (\ref{vdp}). Description of the dynamics as motion of a particle in the potential field $U(y)=\frac{1}{2}y^2$ (upper panels) and corresponding time realization $y(t)$ (lower panels) at $\varepsilon=-0.1$ (a) and $\varepsilon=0.1$ (b). The grey area corresponds to negative values of the dissipation function $\gamma(y)$. }
\label{fig0}
\end{figure}

The described principle is simply illustrated on the example of the Van der Pol self-sustained oscillator\cite{vanderpol1920}, which is described by the following equation:
\begin{equation}
\label{vdp}
\dfrac{d^2 y}{dt^2}+(y^2-\varepsilon)\dfrac{dy}{dt}+y= 0, \\
\end{equation}
where $y$ is the dynamical variable, $\varepsilon$ is the parameter, which determines the dynamics. Autonomous  system (\ref{vdp}) exhibits the regular quiescent or self-oscillatory dynamics, while the non-autonomous model can realize singular solutions \cite{levinson1949}. In terms of motion of a particle in a potential field the system (\ref{vdp}) describes oscillations in the potential $U(y)=\frac{1}{2}y^2+K$ (the constant $K$ is neglected in the following, $K=0$) in the presence of dissipation defined by the factor $\gamma=\gamma(y)=y^2-\varepsilon$. In case $\varepsilon<0$ the dissipation function $\gamma(y)$ is positive  at any values of $y$. Then oscillations $y(t)$ are damped and all trajectories are attracted to a stable equilibrium point in the origin, corresponding to a potential well bottom [Fig. \ref{fig0} (a)]. If $\varepsilon>0$, the function $\gamma(y)$ possesses negative values in the vicinity of a potential well bottom [the grey area in Fig. \ref{fig0} (b)] and the equilibrium point in the origin becomes unstable. The phase point is repelled from the unstable steady state and goes out of the area corresponding to negative $\gamma(y)$. After that the amplitude growth slows down and stops. After transient time stationary periodical self-oscillations are organized. There is energy balance between dissipation and pumping during the period of the self-oscillations.  The same principle of self-oscillation excitation takes place in the Rayleigh \cite{rayleigh1883} self-oscillator, FitzHugh-Nagumo model \cite{fitzhugh1961,nagumo1962} and in other examples of self-sustained oscillators.

Another principle of self-oscillation excitation is described in the current paper. As will be shown below, the self-oscillation excitation can be realized in the system (\ref{general-eq}) in a case of the dissipation factor $\gamma$ being positive at each point of the phase space. This phenomenon becomes possible due to specific features of the potential function assumed to be bivariate, $U=U(y,\frac{dy}{dt})$. In other words, the problem of motions of the point mass in a state-dependent potential field with positive dissipation will be considered. 

At first, the explored issue will be numerically and experimentally studied in the context of a model of a double-well oscillator with state-dependent dissipation and potential, which exhibits hard oscillation excitation. Then a simplified model, Eq. (\ref{general-eq}) with the positive drag coefficient $\gamma>0$, which demonstrates soft self-oscillation excitation, will be proposed. It has to be noted that the used term "self-oscillation excitation under condition of positive dissipation" is correct only in the context of motion of a particle in a potential field. Generally, dissipativity of the considered systems is characterized by the divergence of the phase space flow. The divergence corresponding to the self-oscillatory dynamics in the studied models is positive in some regions of the phase space, while it is negative in other areas. In this point of view the occurrence of self-oscillatory behaviour traced by a stable limit cycle is understandable and logical.

Numerical modelling of the studied systems was carried out by integration using the Heun method \cite{manella2002} with the time step $\Delta t=0.0001$. 
\section{Self-oscillatory motion in a double-well state-dependent potential field in the presence of positive dissipation}
%
\begin{figure}
\centering
\includegraphics[width=0.6\textwidth]{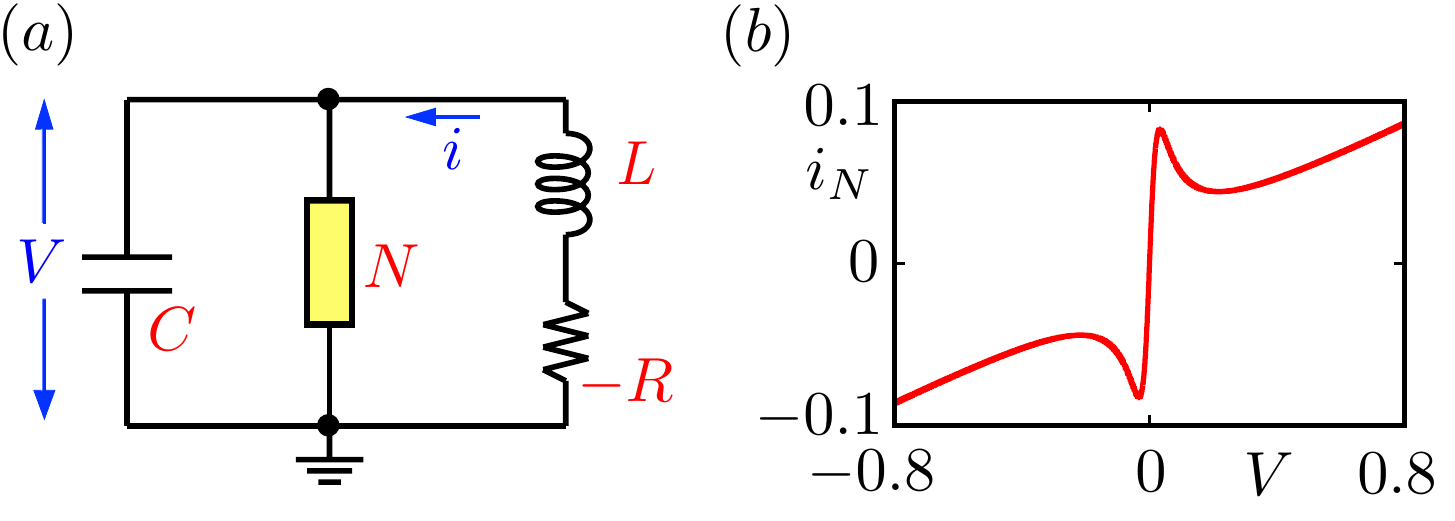} 
\caption{(a) Schematic circuit diagram of double-well oscillator with nonlinear dissipation; (b) Current-voltage characteristic of element N corresponding to parameters $a=200$, $b=0.2$, $g=0.1$.}
\label{fig1}
\end{figure}  
Figure \ref{fig1}~(a) shows the system under study. It is a simplified modification of the bistable self-oscillator, offered in the paper \cite{semenov2016}. Fig.~\ref{fig1}~(a) shows a parallel oscillatory circuit including the negative resistance -R and the nonlinear element N with the current-voltage characteristic $i_{N}(V)=\frac{V}{aV^{2}+b}+gV$ [Fig. \ref{fig1} (b)]. By using the Kirchhoff's current law the following differential equations for the voltage $V$ across the capacitor $C$  and the current $i$ through the inductor $L$ can be derived:
\begin{equation}
\label{eq1}
\left\lbrace
\begin{array}{l}
C\dfrac{dV}{dt'}+i+\dfrac{V}{aV^{2}+b}  + gV = 0, \\
V = L\dfrac{di}{dt'}-Ri. \\
\end{array}
\right.
\end{equation}
In the dimensionless variables $x=V / V_{0}$ and $y=i/ i_{0}$ with $V_{0}= 1$~V, $i_{0}=1$~A and dimensionless time $t~=~[(V_{0}/(i_{0}L)]t'$, Eq.(\ref{eq1}) can be re-written as,
\begin{equation}
\label{initial}
\left\lbrace
\begin{array}{l}
\varepsilon \dot{x} = -y-gx - \dfrac{x}{ax^2+b}, \\
\dot{y}=x+my, \\
\end{array}
\right.
\end{equation}
where $\dot{x}=\frac{dx}{dt}$, $\dot{y}=\frac{dy}{dt}$, the parameter $\varepsilon$ sets separation of slow and fast motions, other parameters are $g,a,b,m~>~0$. 
Eqs. (\ref{initial}) can be written in the "coordinate-velocity" form with the dynamical variables $y$, $v\equiv \dot{y}$:
\begin{equation}
\label{y-v}
\left\lbrace
\begin{array}{l}
\dot{y}=v,\\
\varepsilon \dot{v}=y(mg-1)+v(\varepsilon m-g)\\
+\dfrac{my-v}{a(v-my)^{2}+b}.
\end{array}
\right.
\end{equation}
In the oscillatory form the system (\ref{y-v}) becomes,
\begin{figure}
\centering
\includegraphics[width=0.7\textwidth]{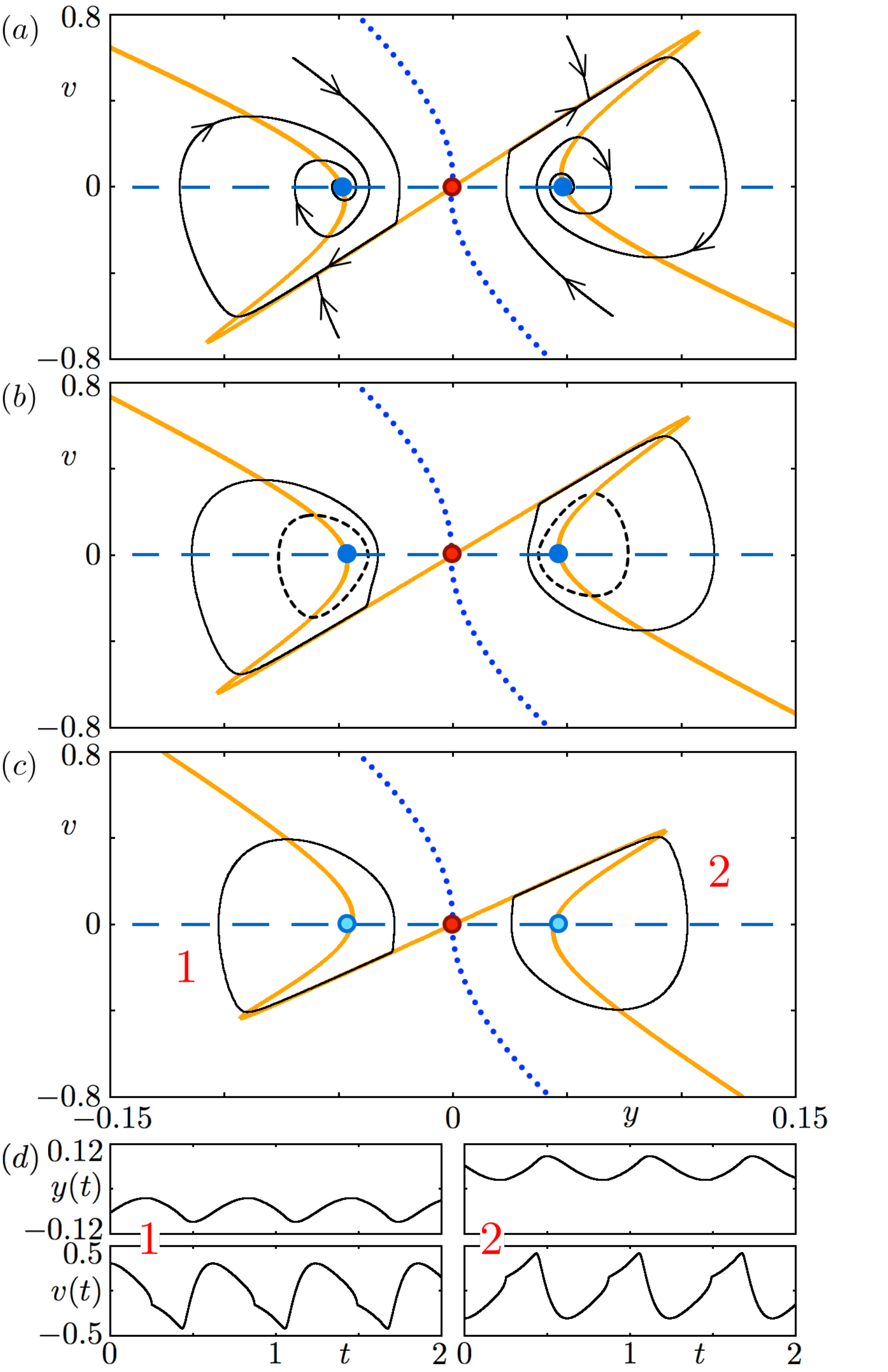} 
\caption{System (\ref{y-v}). Phase space structure corresponding to $m=7$ (a), $m=6.5$ (b), and $m=5$ (c). Left and right equilibrium points are shown by blue circles (stable equilibria are filled by the dark blue colour, the unstable ones have a light blue fill); the saddle point at the origin is coloured in red; the blue dashed line indicates the nullcline $\dot{y} = 0$; the orange solid line shows the nullcline $\dot{v}= 0$; the separatrix of the saddle is shown by the blue dotted line; phase trajectories started from various initial conditions are shown by black arrowed lines; stable limit cycles are shown by black solid lines; unstable limit cycles are shown by black dashed lines. (d) Time traces of state variables in two coexisting self-oscillatory regimes 1 and 2. Parameters are: $\varepsilon=0.005$, $a=200$, $b=0.2$, $g=0.1$.}
\label{fig2}
\end{figure}  
\begin{equation}
\label{oscillatory}
\ddot{y}+q_{1}(y,\dot{y})\dot{y}+q_{2}(y,\dot{y})=0,
\end{equation}
and describes motion in a potential field, where $q_{2} (y,\dot{y})=\dfrac{dU(y,\dot{y})}{dy}$ defines the form of the potential $U(y,\dot{y})$ and $q_{1}(y, \dot{y})$ is the nonlinear dissipation:
\begin{equation}
\label{q1}
q_{1}(y,\dot{y})= \dfrac{1}{\varepsilon}\left(g-m\varepsilon+\dfrac{1}{a(\dot{y}-my)^{2}+b}\right),
\end{equation}
\begin{equation}
\label{q2}
q_{2}(y,\dot{y})= \dfrac{y}{\varepsilon}\left(1-mg-\dfrac{m}{a(\dot{y}-my)^{2}+b}\right). 
\end{equation}
Any attempts to group the terms of Eq. (\ref{oscillatory}) in a different way and to derive different functions $q_{1}(y,\dot{y})$ and $q_{2}(y,\dot{y})$ give rise to appearance of second-order discontinuity in the functions of a potential and dissipation, which contradicts physical nature of the system. For example, one can try to exclude the term $-m/(a(\dot{y}-my)^{2}+b)$ from the function $q_2$ (Eq. (\ref{q2})) and to derive a new function $q_{2}^{*}$ as being: $q_{2}^{*}(y)=\dfrac{y}{\varepsilon}(1-mg)$. Then the new oscillatory form of the system (\ref{y-v}) becomes:
\begin{equation}
\label{oscillatory2}
\ddot{y}+q_{1}^{*}(y,\dot{y})\dot{y}+q_{2}^{*}(y)=0,
\end{equation}
where the function $q_{1}^{*}(y,\dot{y})$ denotes the influence of dissipation:
\begin{equation}
\label{q1-2}
q_{1}^{*}(y,\dot{y})= \dfrac{1}{\varepsilon}\left(g-m\varepsilon+\dfrac{1}{a(\dot{y}-my)^{2}+b}-\dfrac{my}{(a(\dot{y}-my)^{2}+b)\dot{y}}\right).
\end{equation}
The function $q_{1}^{*}(y,\dot{y})$ tends to $-\infty$ at each finite positive value $y$ and an extremely small positive value $\dot{y}\to 0$. On the other hand, the function $q_{1}^{*}(y,\dot{y})$ tends to $+\infty$ at the same value $y$ and an extremely small negative value $\dot{y}\to 0$. Similarly, the function $q_{1}^{*}(y,\dot{y})$ tends to $\pm\infty$ at $\dot{y}\to 0$ and any finite negative values $y$. This is due to appearance of $\dot{y}$ in the denominator of the last term of the formula (\ref{q1-2}). Thus, the function $q_{1}^{*}(y,\dot{y})$ has a manifold of second-order discontinuity points with coordinates ($y\neq0$, $\dot{y}=0$).

Further consideration of the system will be carried out in the variables ($y,v\equiv \dot{y}$) (Eqs. (\ref{y-v}) and (\ref{oscillatory})) with the parameters set to $\varepsilon=0.005$, $a=200$, $b=0.2$, $g=0.1$. The phase space of the system (\ref{y-v}) corresponding to $m=7$ is illustrated in Fig. \ref{fig2} (a). There exist two stable fixed points (stable focuses) and  the saddle point at the origin. Basins of attraction of the stable equilibria are separated by the separatrix [the blue dotted line in Fig.~\ref{fig2}~(a)--(c)] of the saddle. Decreasing of the parameter $m$ leads to the pair saddle-node bifurcation of limit cycles at $m = 6.967$: two pairs of limit cycles (stable and unstable) appear both in the left branch of the phase space and in the right one [Fig.~\ref{fig2}~(b)]. Further decreasing of the parameter $m$ gives rise to the same transformation in the left and right branches of the phase space: unstable limit cycle and stable fixed point collide in the subcritical Andronov-Hopf bifurcation at $m = 5.769$. After that the side equilibria become unstable and two attractors (two stable limit cycles) exist in the phase space [Fig.~\ref{fig2}~(c)]. The self-oscillatory behaviour represents the fast-slow dynamics like in the FitzHugh-Nagumo model. It includes slow motions along the nullcline $\dot{v}=0$ and the fast ones, when the phase point falls down from the nullcline [Fig.~\ref{fig2}~(b)--(d)].

\begin{figure}
\centering
\includegraphics[width=0.7\textwidth]{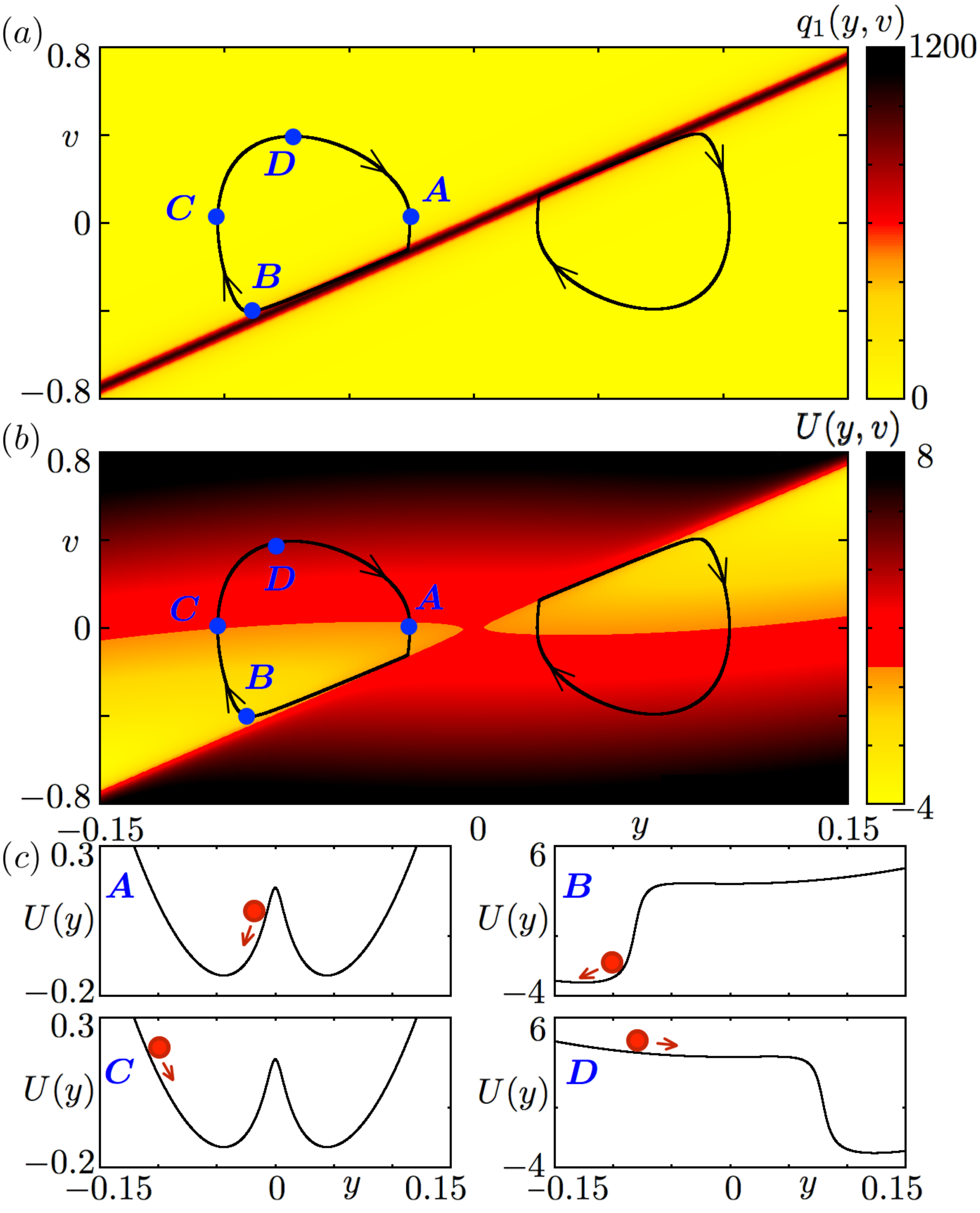} 
\caption{System (\ref{oscillatory}). (a) Dependence of dissipation function, $q_{1}(y,v\equiv \dot{y})$, on system state [Eq. (\ref{q1})]. (b) Potential function, $U(y,v\equiv \dot{y})$ [Eq. (\ref{u})]. (c) Potential function, $U(y)$, corresponding to instantaneous velocity in the points A-D at the limit cycle (is marked by black closed curve in insets (a) and (b)). Direction of point mass motion is schematically shown. Parameters are: $\varepsilon=0.005$, $a=200$, $b=0.2$, $g=0.1$, $m=5$.}
\label{fig3}
\end{figure}  
%

Consideration of the oscillatory form (\ref{oscillatory}) allows to reveal the unusual detail: the drag function $q_{1}(y,v\equiv\dot{y})$ [Eq. (\ref{q1})] is positive at any values of $y$ and $v$ before and after the subcritical Andronov-Hopf bifurcation. In terms of motion in a potential field it means self-oscillation excitation is realized under condition of positive dissipation [Fig. \ref{fig3}(a)]. At first view, this result seems to be false. However, analysis of the function $q_{2}(y,v\equiv\dot{y})$, which is responsible for the potential, allows to explain the contradiction. Using Eq.~(\ref{q2}), one can derive the function of the potential: 
\begin{equation}
\label{u}
\begin{array}{l}
U(y,v)= \dfrac{y^{2}(1-mg)}{2\varepsilon}-\dfrac{1}{2\varepsilon ma} \text{ln} \left((my-v)^2+\dfrac{b}{a} \right) \\
-\dfrac{v}{\varepsilon m\sqrt{ab}}\text{arctg}\left(\dfrac{\sqrt{a}}{\sqrt{b}}(my-v)\right)+K.
\end{array}
\end{equation}
The constant $K$ is neglected in the following, $K=0$. It is important to note that the potential function $U~=~U(y,v)$ is bivariate in the present case [Fig.~\ref{fig3}~(b)]. It means the action of the potential field at any time is defined by the velocity of a particle moving therein. Then nature of the self-sustained oscillations becomes visible. 

Description of the system (\ref{oscillatory}) in the context of motion in a potential field in the control points A-D [Fig. \ref{fig3}] allows to explain the existence of the self-oscillations. Starting from the point A [Fig. \ref{fig3}(c), inset A], the particle falls down to the potential well bottom and moves with acceleration. It leads to deepening of the potential well, which increases the acceleration of the particle. Then the particle reaches the vicinity of the potential well bottom [Fig. \ref{fig3}(c), inset B]. Growth of the velocity stops because of a lower slope of the potential function and the presence of positive dissipation. The velocity decreases after passing of the potential well bottom, which involves inverse transformation of the potential. The potential well gets up as well as the point mass particle. The particle stops at the other side of the potential well and begins to fall down again [Fig. \ref{fig3}(c), inset C] with increasing velocity. At the same time the potential well goes on to go up and to lift the particle up. The particle's velocity reaches the maximal value [Fig. \ref{fig3}(c), inset D] and then the particle slows down its motion. Finally, the point mass returns to the point A. 

\begin{figure}
\centering
\includegraphics[width=0.7\textwidth]{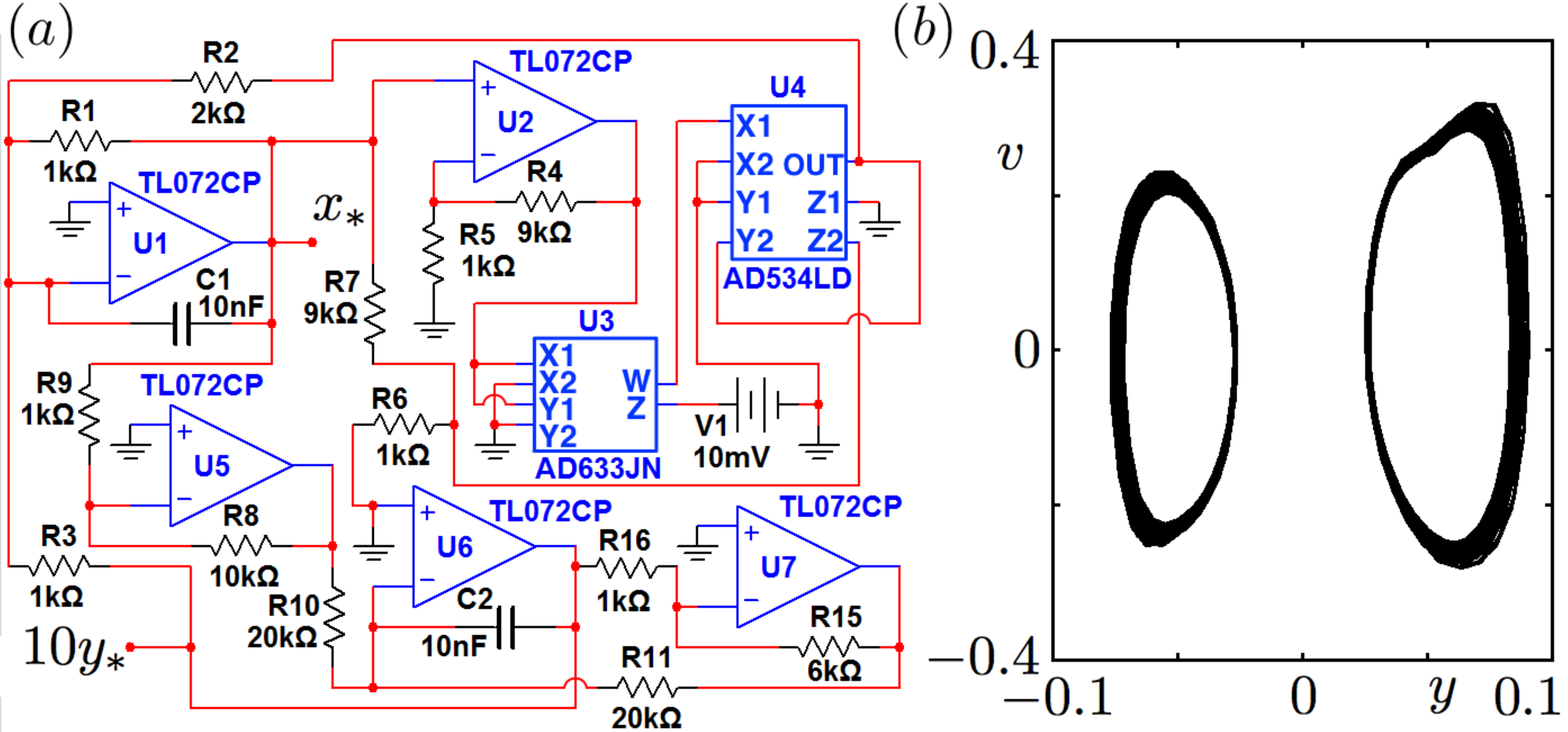} 
\caption{(a) Circuit diagram of setup used in analog experiments [Eq. (\ref{exp})]. (b) Coexistence of two self-oscillatory regimes at $m=6$.}
\label{fig4}
\end{figure}

One can suppose that the described transition to the self-oscillatory dynamics is possible only in the deterministic mathematical models. Sharp change of dissipation [Fig. \ref{fig3}(a)] can give rise to unpredictable results in the presence of fluctuations, which inevitable occur in real systems. In order to verify possibility of self-oscillation excitation in physical experiment, an electronic model of the considered system has been developed. The main problem of electronic model creation was realization of the nonlinearity of the system (\ref{initial}) with high accuracy. Therefore the experimental electronic setup was developed by using principles of analog modelling \cite{luchinsky1998}. The circuit diagram is shown in Fig. \ref{fig4} (a). The experimental facility allows to obtain instantaneous values of the variables $x_{*},y_{*}$ and then $v_{*}=\dot{y}_{*}=x_{*}+my_{*}$. Time series were recorded from corresponding outputs (are marked in Fig.~\ref{fig4} (a)) using an acquisition board (National Instruments NI-PCI 6133). The signals were digitized at the sampling frequency of 200 kHz. 10~s long realizations were used for further offline processing. The circuit in Fig. \ref{fig4}~(a) is described by the following equations:
\begin{equation}
\label{exp}
\left\lbrace
\begin{array}{l}
0.1R_{x}C_{x} \dfrac{dx_{*}}{dt_{*}} = -y_{*}-gx_{*}-\dfrac{x_{*}}{ax_{*}^{2}+b},\\
R_{y}C_{y}\dot{y}_{*}=x_{*}+my_{*}, \\
\end{array}
\right.
\end{equation}
where $C_{x}=C_{y}=10$~nF,  $R_{x}=$1~K$\Omega$ is the resistance at the integrator U1 ($R_{1}=0.5\times R_{2}=R_{3}=R_{x}=1~$K$\Omega$), $R_{y}=20$~K$\Omega$ is the resistance at the integrator U6 ($R_{10}=R_{11}=R_{y}=20$~K$\Omega$). The parameters are $a=200$, $b=0.2$, $g=0.1$, $m=\dfrac{R_{15}}{R_{16}}$. Transition to dimensionless equations (\ref{initial})  with $\varepsilon=\dfrac{0.1R_{x}C_{x}}{R_{y}C_{y}}$ is then carried out by 
substitution $t=t_{*}/\tau_0$, $x=x_{*}/V_{0}$, $y=y_{*}/V_{0}$, where $\tau_0=R_{y}C_{y}=2$~ms is the circuit's time constant  and $V_{0}=1$~V.

%

Electronic experiments have shown stability of the self-oscilatory dynamics in the system (\ref{exp}) [Fig. \ref{fig4} (b)]. There exists quantitative difference between two coexisting regimes of the self-sustained oscillations. Nevertheless, they are stable despite fluctuations and internal unique features of the experimental setup.

\section{Simplified single-well model}
There are additional aspects of system's (\ref{y-v}) dynamics, which make it more complicated, but have no principal impact on the self-oscillation excitation mechanism: the bistability, the existence of slow-fast dynamics and nonlinear state-dependent dissipation. In order to make the phenomenon more evident, the simplified mathematical model has to be proposed for consideration. The offered system must be presented in the form (\ref{general-eq}) with the fixed drag coefficient $\gamma=\text{const}>0$ and a nonlinear term, which is responsible for the potential function and cannot be grouped with the term $\gamma \dot{y}$ without appearance of discontinuity in the dissipation function. The system satisfying all requirements is the following:
\begin{equation}
\label{simple}
\ddot{y}+\gamma\dot{y}+y-\dfrac{1}{(y-\dot{y})^{2} + 1}=0.
\end{equation}
\begin{figure}[t]
\begin{center}
\includegraphics[width=0.9\columnwidth]{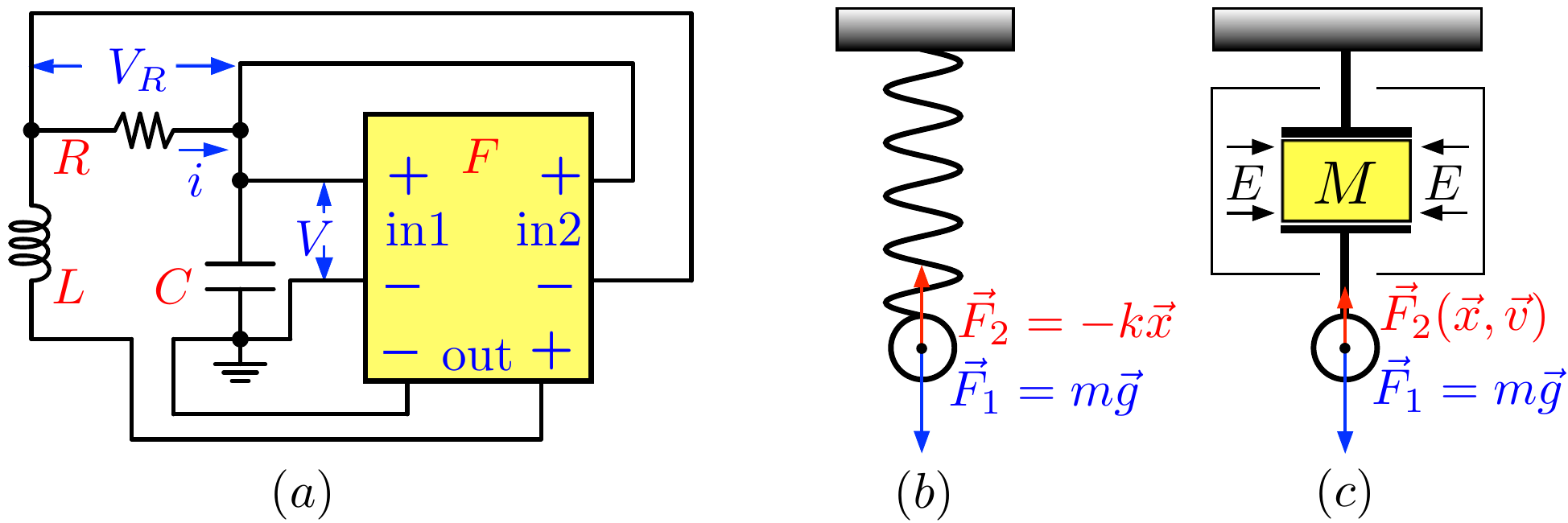}
\end{center}
\caption{(a) Schematic circuit diagram of single-well self-oscillator (Eq.~(\ref{simple})). (b) Model of linear spring pendulum. (c) Modified spring-like pendulum based on elastic medium block.}
\label{fig5}                             
\end{figure}
The system (\ref{simple}) describes, for instance, a schematic electronic circuit depicted in Fig. \ref{fig5} (a). It is a series-oscillatory LCR-circuit driven by a nonlinear feedback. The nonlinear feedback is realized by the nonlinear converter $F$. The converter $F$ has two inputs $V$ (the voltage across the capacitor $C$) and $V_{R}$ (the voltage across the resistor $R$) with zero input current. In that case the same current $i$ passes through the inductor $L$, the resistor $R$, the capacitor $C$ and then $V_{R}=iR$. The converter $F$ has the output voltage as being: $V_{out}=\frac{1}{(V-kV_{R})^2+1}=\frac{V_g}{(V-m i)^2+1}$, where $m=kR$. By using the Kirchhoff's voltage law the following differential equations for the voltage $V$ across the capacitor $C$ can be derived:
\begin{equation}
\label{simple-lcr}
CL\dfrac{d^2V}{dt'^2}+RC\dfrac{dV}{dt'}+V=\dfrac{1}{\left(V-m \frac{dV}{dt'}\right)^2+1}.
\end{equation}
Equation (\ref{simple-lcr}) can be transformed into the dimensionless form (\ref{simple}) with variable $y=V / v_{0}$ and  time $t=t'/m$, where $v_{0}= 1$~V.

One can imagine mechanical realization of the system (\ref{simple}). One of the simplest mechanical oscillators is a spring-based pendulum [Fig. \ref{fig5} (b)]. In a linear pendulum case a spring-suspended solid is affected by the force of gravity, $\vec{F}_{1}=m\vec{g}$, and the elastic force, $\vec{F}_{2}$, being proportional to displacement $\vec{x}$ from equilibrium: $\vec{F}_{2}=-k\vec{x}$. Taking into account the influence of air resistance assumed to be proportional to the velocity of motion $\vec{F}_{3}=-\gamma \vec{v}$, one can derive an equation of motion in the vectorial form (Newtown's second law): $m\vec{a}=\vec{F}_{1}+\vec{F}_{2}+\vec{F}_{3}$. Then the scalar form of the equation is: $m\ddot{x}+\gamma \dot{x}+kx=0$. Generally, elastic properties of objects can be more complex (see for example a model of hair bundles with negative stiffness \cite{martin2000,martin2003}). Moreover, the spring can be changed to a complex device [Fig. \ref{fig5} (c)] including elastic medium (the medium $M$ in Fig. \ref{fig5} (c)) and a source of energy (is marked by the source E in Fig. \ref{fig5} (c)). Then one can assume that the elastic force nonlinearly depends both on the displacement $x$ and an instantaneous value of the velocity $\dot{x}$ according to the formula: $F_{2}(x,\dot{x})=-x+\frac{1}{(x-\dot{x})^2+1}$. Then the equation of motion becomes:
\begin{equation}
\label{simple-mechanical}
m \ddot{x} + \gamma \dot{x} + x - \frac{1}{\left(x-\dot{x}\right)^2+1}= 0, \\
\end{equation}
which coincides with  Eq. (\ref{simple}) at $m=1$. Here the action of nonlinear feedback is hidden in the structure of a complex elastic device including the medium $M$.

\begin{figure}
\centering
\includegraphics[width=0.7\textwidth]{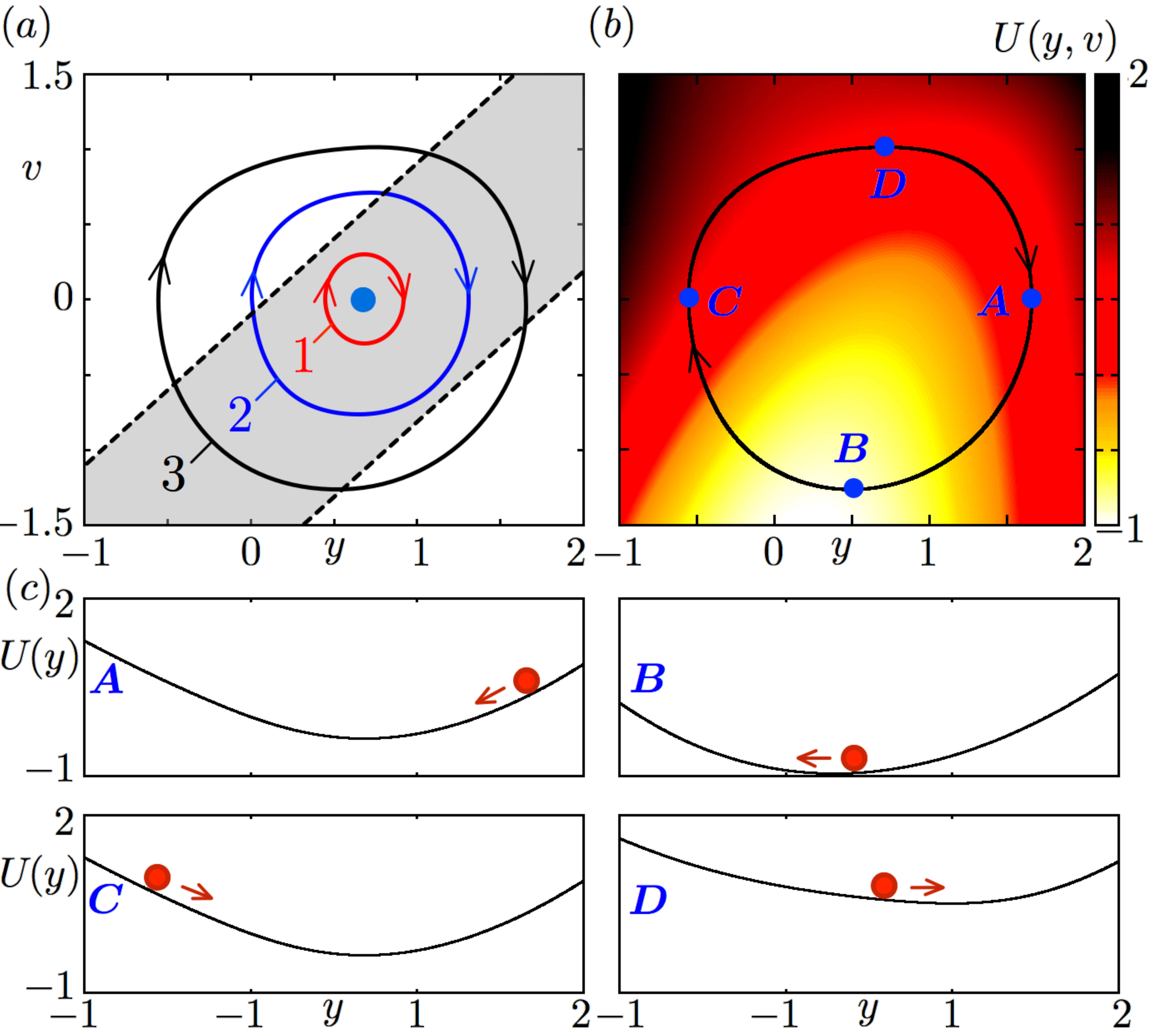} 
\caption{System (\ref{simple}) in variables $y$, $v\equiv \dot{y}$. (a) Evolution of dynamics caused by dissipation parameter $\gamma$ decreasing: steady state regime at $\gamma=0.7$ (the blue circle is the stable focus), self-oscillatory regime at $\gamma=0.6$ (the red stable limit cycle 1), $\gamma=0.4$ (the blue stable limit cycle 2), and $\gamma=0.2$ (the black limit cycle 3). Grey background shows the area of positive divergence (Eq. (\ref{divergence})) at $\gamma=0.2$, white background indicates regions of negative values of the divergence at $\gamma=0.2$. (b) Potential function, $U(y,v)$ [Eq. (\ref{u-s})] and the stable limit cycle 3 in inset (a). (c) Potential function, $U(y)$, corresponding to instantaneous velocity in points A-D at limit cycle in inset (b). Direction of point mass motion is schematically shown.}
\label{fig6}
\end{figure}

Equation (\ref{simple}) describes motions of a particle in the potential field $U(y,v\equiv\dot{y})$ as being:
\begin{equation}
\label{u-s}
\begin{array}{l}
U(y,v)= \dfrac{y^{2}}{2} - \text{arctg}(y-v)+K,
\end{array}
\end{equation}
in the presence of positive dissipation. The constant $K$ is neglected, $K=0$. As in the previous system [Eq.~(\ref{oscillatory})], the form of the potential at any time is determined by an instantaneous value of the point mass velocity, $v$. In the presence of the high dissipation factor $\gamma$ an attractor in the phase space of the system (\ref{simple}) is a stable equilibrium point with the coordinates $y_{0}=\sqrt [3]{\frac{1}{2}+\sqrt{\frac{31}{108}}}+\sqrt [3]{\frac{1}{2}-\sqrt{\frac{31}{108}}}\approx 0.682$, $v_{0}=0$ [the blue circle in Fig.~\ref{fig6} (a)].  Decreasing of the drag coefficient $\gamma$ leads to the loss of stability of the fixed point, and a stable limit cycle appears in the vicinity of the unstable point of equilibrium. The Andronov-Hopf scenario of soft self-sustained oscillation excitation is realized at $\gamma_{0}=\frac{2y_{0}}{(y_{0}^{2}+1)^2}\approx 0.6355$. Further decreasing of the parameter $\gamma$ leads to extension of the limit cycle [the limit cycles 1-3 in Fig.~\ref{fig6}~(a)]. The self-oscillation excitation principle is based on features of a state-dependent potential [Fig. \ref{fig6} (b),(c)] similarly to the system (\ref{oscillatory}). Depending on the velocity of the particle, the potential moves it up and down and changes position of the local minimum of the potential well function [Fig.~\ref{fig6}~(c)].

The system (\ref{simple}) exhibits self-oscillation excitation at the positive coefficient of dissipation, $\gamma$. At the same time, the divergence of the phase space flow determined by the formula
\begin{equation}
\label{divergence}
\text{div}G = -\gamma + \dfrac{2(y-\dot{y})}{((y-\dot{y})^2+1)^2}
\end{equation}
is an alternating quantity (see Fig. \ref{fig6} (a)). The existence of a positive divergence area in the phase space at positive drag coefficient $\gamma$ means that the system includes a source of energy, the action of which is reflected in peculiarities of a potential function. The same principle holds true in the system (\ref{y-v}).

%

\section{Conclusions} 
Classical representation of the self-oscillatory dynamics from the perspective of motion of a particle in a potential field involves the presence of a changeless potential and state-dependent dissipation, which possesses negative and positive values and is responsible for the existence of the self-sustained oscillations. Another configuration of self-oscillatory motion in a potential field is proposed in the present paper. It implies positive dissipation factor and a state-dependent potential, which is responsible for self-oscillation excitation and energy dissipation-supply balance in the self-oscillatory regime. In that case both steady state instability and amplitude limitation are dictated by the potential function in the studied systems. The described effect corresponds to mutual interaction between the point mass particle and the potential field. The potential determines dynamics of the particle, but it changes depending on the particle's velocity. Despite special aspects of interpretation as motion of a particle in a potential field, self-oscillation excitation in the considered systems corresponds to well-studied bifurcation mechanisms associated with the Andronov-Hopf bifurcation. 

It is important to note that the proposed interpretation of self-oscillatory motion in a state-dependent potential field is discussed in the context of autonomous systems. It allows to distinguish between the described velocity-dependent configuration of the potential and the time-dependent potential well used for description of parametric resonance, stochastic resonance and other effects observed in non-autonomous systems.  

There are known examples of self-oscillators described by equations similar to (\ref{general-eq}) with a positive coefficient of dissipation or a state-dependent function of dissipation being positive at any point of the phase space (see for example papers devoted to thermodynamic self-oscillators \cite{finnie1963,finnie1963-2,alicki2017}). The presented in the current paper results allow to expand a group of such oscillators and provide a method of description of their properties in an unified manner.

\section*{Acknowledgments}
This work was supported by the Russian Ministry of Education and Science (project code 3.8616.2017/8.9).




\end{document}